\documentclass[12pt, draftclsnofoot, jounal, a4paper, onecolumn]{IEEEtran}
\usepackage[dvips]{graphicx}
\usepackage{times}
\usepackage{color}
\usepackage{mathrsfs}
\usepackage{amsmath}
\usepackage{amssymb}
\usepackage{cite}
\usepackage{subfigure}
\begin{document}
%
\title{End-to-End Outage Minimization in
OFDM Based Linear Relay Networks}
\author{\authorblockN{Xiaolu Zhang, Meixia Tao, Wenhua Jiao and
Chun~Sum~Ng}\thanks{X. Zhang and C. S. Ng are with the Department of
Electrical and Computer Engineering, National University of
Singapore, Singapore 117576 (e-mail: zhangxiaolu@nus.edu.sg;
elengcs@nus.edu.sg). Meixia Tao is with Dept. of Electronic
Engineering, Shanghai Jiao Tong University, Shanghai, China 200240
(e-mail: mxtao@sjtu.edu.cn). W. Jiao is with Bell labs Research
China, Alcatel-Lucent, Beijing, P.R. China 100080 (e-mail:
wjiao@alcatel-lucent.com).}}

\maketitle
\newtheorem{lemma}{Lemma}
\newtheorem{theorem}{Theorem}
\newtheorem{proposition}{Proposition}
\begin{abstract}

Multi-hop relaying is an economically efficient architecture for
coverage extension and throughput enhancement in future wireless
networks. OFDM, on the other hand, is a spectrally efficient
physical layer modulation technique for broadband transmission. As a
natural consequence of combining OFDM with multi-hop relaying, the
allocation of per-hop subcarrier power and per-hop transmission time
is crucial in optimizing the network performance. This paper is
concerned with the end-to-end information outage in an OFDM based
linear relay network. Our goal is to find an optimal power and time
adaptation policy to minimize the outage probability under a
long-term total power constraint. We solve the problem in two steps.
First, for any given channel realization, we derive the minimum
short-term power required to meet a target transmission rate. We
show that it can be obtained through two nested bisection loops. To
reduce computational complexity and signalling overhead, we also
propose a sub-optimal algorithm. In the second step, we determine a
power threshold to control the transmission on-off so that the
long-term total power constraint is satisfied. Numerical examples
are provided to illustrate the performance of the proposed power and
time adaptation schemes with respect to other resource adaptation
schemes.

\end{abstract}
\IEEEpeerreviewmaketitle

\renewcommand{\vec}[1]{\mbox{\boldmath$#1$}}
\section{Introduction}

Relay networks in the form of point-to-multipoint based tree-type or
multipoint-to-multipoint mesh-type architectures are a promising
network topology in future wireless systems. The basic concept of
relaying is to allow a source node to communicate with a destination
node under the help of a single or multiple relay nodes. It has been
shown that relaying can bring a wireless network various benefits
including coverage extension, throughput and system capacity
enhancement. Recently, multi-hop relaying has been widely adopted in
wireless networks such as next generation cellular networks,
broadband wireless metropolitan area networks and wireless local
area networks.
On the other hand, orthogonal frequency division multiplexing (OFDM)
is an efficient physical layer modulation technique for broadband
wireless transmission. It divides the broadband wireless channel
into a set of orthogonal narrowband subcarriers and hence eliminates
the inter-symbol interference.
OFDM is one of the dominating transmission techniques in many
wireless systems, e.g., IEEE 802.16 (WiMax), EV-DO Revision C and
the Long-Term-Evolution (LTE) of UMTS. The combination of OFDM and
multi-hop relaying has received a lot of attention recently. For
example, this OFDM-based relay architecture has been proposed by the
current wireless standard IEEE 802.16j \cite{PabstWSH04j}. The
complexity of relay station is expected to be much less than the one
of legacy IEEE 802.16 base stations, thereby reducing infrastructure
deployment cost and improving the economic viability of IEEE 802.16
systems \cite{WiMax16j}.
%
%
%

In this work, we are interested in an OFDM-based linear relay
network. The so-called linear relay network consists of
one-dimensional chain of nodes, including a source node, a
destination node and several intermediate relay nodes. It can be
viewed as an important special case of relay networks where only a
single route is active. As a natural consequence of multi-hop
relaying and OFDM transmission, the allocation of per-hop subcarrier
power and per-hop transmission time is crucial in optimizing the
end-to-end network performance.
%
%
%

Previous work on resource allocation for relay networks is found in
\cite{YaoCG05j,SikoraLHC06j,OymanLS06j,RadunovicB03c,LiL06j,DaiGC07}.
Yao \emph{et al.} in \cite{YaoCG05j} considered a classic three-node
network (a source node, a destination node, and a relay node) and
compared the energy required for transmitting one information bit in
different relay protocols. Authors in \cite{SikoraLHC06j} and
\cite{RadunovicB03c} studied efficient scheduling and routing
schemes in one-dimensional multi-hop wireless networks.  It is
assumed in all these works
\cite{SikoraLHC06j,RadunovicB03c,YaoCG05j} that the point-to-point
links are frequency-flat fading channels and the system has a fixed
short-term power constraint. In \cite{OymanLS06j}, Oyman \emph{et
al.} summarized the end-to-end capacity results of a multi-hop relay
network under fixed-rate and rate-adaptive relaying strategies, and
further illustrated the merits of multi-hop relaying in cellular
mesh networks.
%
Authors in \cite{ZhangJT08c} studied the per-hop transmission time
and subcarrier power allocation problem in the OFDM based linear
multi-hop relay network to maximize the end-to-end average
transmission rate under a long-term total power constraint. However,
the end-to-end average rate can only be obtained at the expense of
large delay.


For many real-time services, one has to maintain a target
transmission rate and avoid service outage in most of fading
condition through adaptive resource allocation. An outage is an
event that the actual transmission rate is below a prescribed
transmission rate (\cite{CaireTB99j} and \cite{LiG00j2}). Outage
probability can be viewed as the fraction of time that a codeword
is decoded wrongly. For any finite average power constraint,
transmission outage may be inevitable over fading channels.
However, one can minimize the outage probability through adaptive
power control \cite{CaireTB99j}.
%
%
In a relay network where no data is allowed to accumulate at any of
relay nodes, an end-to-end outage is the event that there exists a
hop over which the transmission rate is lower than the target rate.

The goal of this paper is to investigate the optimal per-hop power
and time control to minimize the end-to-end outage probability in an
OFDM linear relay network under an average transmission power
constraint. At first, we derive the minimum short-term power
required to meet a target transmission rate for any given channel
realization. The resulting power and time allocation can be obtained
through a Two-nested Binary Search (TBS) which is conducted in a
central controller with the knowledge of channel state information
(CSI) on all subcarriers and over all hops. Such algorithm gives a
theoretical performance limit of linear relay networks, but is
computationally intense. Moreover, it requires significant
signalling overhead and channel feedback between network nodes and
the central controller. For this reason, an Iterative Algorithm of
Sub-optimal power and time allocation (IAS) is proposed. The
required information for signalling exchange only involves the
number of active subcarriers on each hop and the geometric mean and
harmonic mean of channel gains averaged over the active subcarriers.
This sub-optimal allocation algorithm suggests prolonging the
transmission time for the hop with low geometric mean of channel
gains while lowering the transmission power for the hop with low
harmonic mean. After obtaining the minimum power required to support
the target rate for a given channel realization, we then compare it
with a threshold. The transmission will be cut off if the required
minimum total power exceeds the threshold. The threshold takes the
value so that the long-term total power constraint is satisfied.
Numerical results show that a significant power saving can be
achieved by the proposed optimal power and time allocation compared
with the uniform power and time allocation under the same end-to-end
outage probability. In addition, the proposed sub-optimal power and
time allocation serves as a good approximation to the optimal
solution when the target rate is sufficiently high. The optimal
number of hops in the sense of requiring minimum power at different
target rates is also shown numerically.

The remainder of this paper is organized as follows. In Section
\ref{sec_system_model}, the system model and problem formulation are
presented. The optimal and sub-optimal resource allocation
algorithms to minimize the end-to-end outage probability under an
average total power constraint are proposed in Section
\ref{sec_APTA}. Numerical results are given in Section
\ref{sec_numerical}. Finally we conclude this paper in Section
\ref{sec_conclusions}.

\section{System model, end-to-end rate and outage probability}\label{sec_system_model}
\subsection{System Model}
\begin{figure}
\centering
\includegraphics[width=5in]{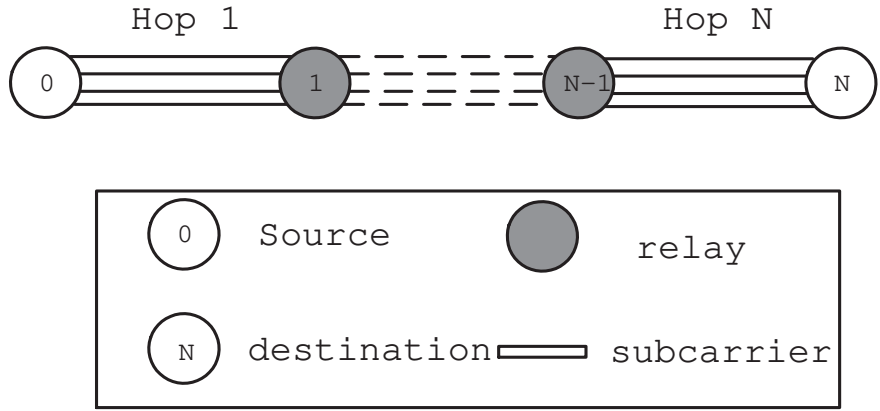}
\caption{Illustration of linear relay networks}
\label{fig_linear_network}
\end{figure}

Consider a wireless linear relay network shown in Fig.
\ref{fig_linear_network}. The source node $R_0$ communicates with
destination $R_N$ by routing its data through $N-1$ intermediate
relay nodes $R_n~(n = 1, \ldots, N-1)$. The hop between node
$R_{n-1}$ and $R_n$ is indexed by $n$, and the set of hops is
denoted by $\mathcal{N}$. We focus on time-division based relaying.
The transmission time is divided into frames of multiple time slots.
Within every time frame, the transmission over each hop takes place
at the assigned time slots. In general, frequency reuse can be
applied so that more than one hop can be transmitting at a same time
slot. However, due to interference issue, it will increase decoding
complexity as well as decoding delay \cite{SikoraLHC06j}. Thus, in
this work we do not pursue the frequency reuse. In each time frame,
the message from the source is sequentially relayed at each hop
using decode-and-forward protocol \cite{LanemanTW04j}. Each relay
decodes the message forwarded by the previous node, re-encodes it,
and then transmits it to the next receiver. The channel for each hop
is assumed to be a block fading Gaussian channel, and the channel
coefficients remain constant during the entire time frame but change
randomly from one frame to another. Over each hop, OFDM with $K$
subcarriers is used as the physical layer modulation scheme. We
denote the set of subcarriers by $\mathcal{K}$. The channel gain on
subcarrier $k$ over hop $n$ in a time frame is denoted as $g_{k,n}$
and it is independent for different $n$.

\subsection{End-to-end rate and outage probability}

Suppose each time frame contains $T$ OFDM symbols and hop $n$ is
scheduled to transmit over $T_n$ OFDM symbols with $T_n$ satisfying
$\sum_{n\in \mathcal{N}}{T_n}=T$.
Then we define the time-sharing fraction as $\rho_n\triangleq
\frac{T_n}{T}$. It is assumed that $T$ is large enough so that
$\rho_n$ can take an arbitrary value between 0 and $1$.
Let $p_{k,n}$ denote the transmission power on subcarrier $k$ over
hop $n$. It is subject to a long-term total power constraint $P$,
given by
\begin{equation}\label{eqn:power_constraint}
    \mathbb{E}\left[\sum_{n\in \mathcal{N}}{\rho_n \sum_{k\in\mathcal{K}}{p_{k,n}}}
    \right] \leq P.
\end{equation}
%
%
%
Then the instantaneous transmission rate in Nat/OFDM symbol in a
time frame achieved over hop $n$ can be written as
%
%
%
%
\begin{equation}\label{eqn_real_transmission_rate}
r_n=\rho_n\sum_{k\in\mathcal{K}}
\ln\left(1+\frac{g_{k,n}p_{k,n}}{\Gamma N_0}\right),~~\forall
n\in\mathcal{N},
\end{equation}
where $N_0$ is the noise power, and $\Gamma$ is the signal-to-noise
ratio (SNR) gap related to a given bit-error-rate (BER) constraint
\cite{QiuC99j}.
%
%
For notation brevity, in the remaining part of this paper, we
redefine $g_{k,n}$ as $g_{k,n} := g_{k,n}/(\Gamma N_0)$.
Under the assumption that no data is allowed to accumulate at any
relay nodes (also called ``information-continuous relaying" in
\cite{OymanLS06j}), the total number of bits received at the
destination node at the end of time frame, $B$, is the minimum of
the number of bits transmitted over each hop, $B_n$, where $B_n=r_n
T_n$.
Thus, the end-to-end transmission rate $r$ can be defined as $r=
\min_{n\in\mathcal{N}} r_n$. In the following we introduce the
end-to-end rates under different resource adaptation policies.
%

\emph{\textbf{Uniform power and time allocation (UPT)}}: When each
transmitting node has no CSI, or does not exploit CSI due to high
signalling overhead, the transmission scheme is independent of the
CSI and both the time and power are uniformly allocated. Hence, the
end-to-end rate can be achieved as
\begin{equation}\label{eqn:r_upt}
r(\mathbf{g},P)=\frac{1}{N} \min_{n\in\mathcal{N}}
\sum_{k\in\mathcal{K}} \ln \left(1+\frac{g_{k,n} P}{K}\right),
\end{equation}
where $\mathbf{g} = \{g_{k,n}, k\in\mathcal{K}, n\in \mathcal{N}\}$.
 As can be seen, the end-to-end rate is limited by the hop
with the worst channel condition.

%
%

 \emph{\textbf{Fixed power and adaptive time allocation
(FPAT)}:} When the transmitters have CSI to some extend (not
necessarily global CSI), each node can perform rate adaptation to
avoid the situation that the ill-conditioned hop become the
bottleneck of the whole link. We assume that the transmission power
on each subcarrer over each hop keeps unchanged, and rate-adaptation
is performed by adjusting time-sharing fraction such that $r_n=r_i,
\forall i\neq n\in\mathcal{N}$.
In this scenario, the maximum end-to-end transmission rate is given
by \cite{OymanS06c}
\begin{eqnarray} \label{eqn:r_fpat}
r(\mathbf{g},\vec{\rho},P)&=&\left(\sum_{n\in\mathcal{N}}\frac{1}{\sum_{k\in\mathcal{K}}
\ln(1+g_{k,n}P/K)}\right)^{-1}.
\end{eqnarray}
This rate is achieved by assigning the time sharing fractions
$\vec{\rho} = \{\rho_i, i\in \mathcal{N}\}$ to be
\begin{equation}\label{eqn_sub-optimal1_rho}
\rho_i(\mathbf{g})=\frac{\prod_{n\neq i}
\sum_{k\in\mathcal{K}}\left[\ln\left(1+g_{k,n}P/K\right)\right]^+}{\sum_{n\in\mathcal{N}}\prod_{j\neq
n}
\sum_{k\in\mathcal{K}}\left[\ln\left(1+g_{k,j}P/K\right)\right]^+},
~i\in\mathcal{N}.
\end{equation}%
%
%
By comparing (\ref{eqn:r_fpat}) with (\ref{eqn:r_upt}), it is found
that the end-to-end rate is increased by adaptive time allocation.
This is because the harmonic mean of a set of nonnegative values is
always greater than or equal to the minimum value.
%
To implement FPTA, the central controller only needs each hop to
feedback the value of $\sum_{k\in\mathcal{K}} \ln(1+g_{k,n}P/K)$
instead of collecting the global CSI $\mathbf{g}$.

%
%

\emph{\textbf{Adaptive power and fixed time allocation (APFT)}}: In
this case the time-sharing fractions are assumed to be fixed and
equal to each other, but the transmission power can be adjusted
adaptively. Hence, the conditional end-to-end rate  for a given
power value set $\mathbf{p}=\{p_{k,n}, k\in\mathcal{K},
n\in\mathcal{N}\}$ is expressed as
\begin{eqnarray} \label{eqn:r_apft}
r(\mathbf{g},
\mathbf{p})&=&\frac{1}{N}\min_{n\in\mathcal{N}}\left[\sum_{k\in\mathcal{K}}
\ln\left(1+ g_{k,n} p_{k,n} \right)\right].
\end{eqnarray}
The set of power values $\mathbf{p}$ depends on the global CSI
$\mathbf{g}$ and the target end-to-end rate.
%
%
%
%
%

\emph{\textbf{Adaptive power and time allocation (APT)}}: We shall
now focus on the scenario of interest, where both transmission power
over each subcarrier and each hop and time over each hop are allowed
to be dynamically allocated. We assume that at the start of each
time frame, the global CSI is perfectly known at a central
controller, which could be embedded in the source node. The
instantaneous end-to-end rate for given power values $\mathbf{p}$
and time-sharing fractions $\vec\rho$ is expressed as
\begin{eqnarray} \label{eqn:r_apt}
r(\mathbf{g}, \vec\rho,
\mathbf{p})&=&\min_{n\in\mathcal{N}}\left[\rho_n\sum_{k\in\mathcal{K}}
\ln\left(1+ g_{k,n} p_{k,n} \right)\right].
\end{eqnarray}

Let $\vec\rho$ satisfy the time constraint $\sum_{n\in{\mathcal
N}}{\rho_n}=1$ and $\mathbf{p}$ satisfy the long-term power
constraint (\ref{eqn:power_constraint}). The end-to-end information
outage probability evaluated at target rate $R$ for APT can be
expressed as
\begin{equation}\label{eqn:outage_APT}
    P_{\rm{APT}}^{out}(R, P) = P(r(\mathbf{g}, \vec\rho,\mathbf{p}) < R).
\end{equation}
Our goal is to minimize $P_{\rm{APT}}^{out}(R, P)$ with respect to
the power and time adaption $\{\mathbf{p(\mathbf{g})},
\vec\rho(\mathbf{g})\}$.
%
%
%
Namely,
\begin{eqnarray}\label{eqn_min_outage}
\textbf{\mbox{P1:}}~\min_{\{\rho_n, p_{k,n}\}}&& P_{\rm{APT}}^{out}(R, P) \\
\mbox{s.t.}&&\mathbb{E}\left[\sum_{n\in\mathcal{N}}\rho_n(\mathbf{g})\left(\sum_{k\in\mathcal{K}}p_{k,n}(\mathbf{g})\right)\right]\leq
P\nonumber\\
&&\sum_{n\in\mathcal{N}} \rho_n(\mathbf{g})=1.
\end{eqnarray}
%

The next section is dedicated to solving the problem P1. As it will
be clear later, APFT can be viewed as a special case of APT by
fixing $\rho_n = 1/N$, $\forall n$ and hence the minimization of its
outage probability can be solved similarly.

\section{Adaptive power and time allocation (APT)}\label{sec_APTA}
 The minimum outage
probability problem \textbf{P1} defined in the previous section can
be generally solved in two steps as proposed in \cite{CaireTB99j}.
First, for each global channel state $\mathbf{g}$, the short-term
minimum total power $p_{\min}(\mathbf{g})$ required to guarantee the
target end-to-end transmission rate $R$ is to be determined. The
second step then determines a threshold to control the transmission
on-off subject to a long-term power constraint.

\subsection{Short-Term Power Minimization}\label{subsec_A}
In this subsection, we shall find the optimal time sharing fraction
$\rho_n^*~(\forall n\in\mathcal{N})$ and the optimal power
allocation $p_{k,n}^*~(\forall n\in\mathcal{N}, k\in\mathcal{K})$ to
minimize the short-term total power needed to achieve a target
end-to-end transmission rate $R$. Then a sub-optimal algorithm with
reduced complexity is developed. The sub-optimal one has a
closed-form expression from which a few attractive properties
regarding time and power allocation can be observed. Comparison on
average powers and computational complexity between the optimal and
sub-optimal algorithms is also given.

\subsubsection{Optimal power and time
allocation}\label{subsubsec_opt}
 The optimal power and time
allocation problem to minimize short-term total power can be
formulated as
\begin{eqnarray}\label{eqn_short_term_power_min}
\textbf{\mbox{P2:}}~p_{\min}(\mathbf{g})=\min_{\{\rho_n,~p_{k,n}\}}
~\sum_{n\in\mathcal{N}}\rho_n(\mathbf{g})\left(\sum_{k\in\mathcal{K}}p_{k,n}(\mathbf{g})\right)\\
\label{eqn_const_target rate}~\mbox{s.t.}~ r(\mathbf{g},\vec{\rho},\mathbf{p})\geq R\\
~\sum_{n\in\mathcal{N}} \rho_n =1.\nonumber
\end{eqnarray}
Unfortunately, the function $r(\mathbf{g},\vec{\rho},\mathbf{p})$
defined in (\ref{eqn:r_apt}) is not concave in $\vec{\rho}$ and
$\mathbf{p}$. As a result, the problem \textbf{P2} is not convex. To
make the problem \textbf{P2} more tractable, we introduce a new
variable $s_{k,n}$ defined as $s_{k,n}:=\rho_{n} p_{k,n}$. This new
variable can be viewed as the actual amount of energy consumed by
hop $n$ on subcarrier $k$ in a time frame interval.
In addition, it follows from (\ref{eqn:r_apt}) that constraint
(\ref{eqn_const_target rate}) can be rewritten as $N$
sub-constraints. By doing these, problem \textbf{P2} is transformed
into a new problem with optimization variables $\rho_n~(\forall
n\in\mathcal{N})$ and $s_{k,n}~(\forall
n\in\mathcal{N},~k\in\mathcal{K})$:
\begin{eqnarray}\label{eqn_short_term_power_min_2}
\textbf{\mbox{P3:}}~\min_{\{\rho_n,~s_{k,n}\}}
&&\sum_{n\in\mathcal{N}}\sum_{k\in\mathcal{K}}s_{k,n}\\
\label{eqn_const_target rate_2}\mbox{s.t.}&& \rho_n \sum_{k\in\mathcal{K}} \ln\left(1+\frac{g_{k,n} s_{k,n}}{\rho_n}\right) \geq R,~~\forall n\in\mathcal{N}\\
\label{eqn_const_rho}&&\sum_{n\in\mathcal{N}} \rho_n =1.
\end{eqnarray}
Since its Hessian matrix is negative semidefinite, the function
$\rho_n \ln(1+ g_{k,n} s_{k,n}/\rho_n)$ is concave in $\rho_n$ and
$s_{k,n}$. Therefore, problem \textbf{P3} is a convex optimization
problem and there exists a unique optimal solution. To observe the
structure of the optimal solution, we write the Lagrangian of
Problem \textbf{P3} as follows:
\begin{eqnarray}\label{eqn_Lagrangian}
J(\{\rho_n\},\{s_{k,n}\},\{\lambda_n\},\beta)=\sum_{n\in\mathcal{N}}
\sum_{k\in\mathcal{K}} s_{k,n}
+\beta\left(\sum_{n\in\mathcal{N}}\rho_n-1\right)+\nonumber\\
~~~~~~~\sum_{n\in\mathcal{N}} \lambda_n\left[R-\rho_n
\sum_{k\in\mathcal{K}}\ln\left(1+\frac{g_{k,n}
s_{k,n}}{\rho_n}\right)\right]
\end{eqnarray}
where $\lambda_n\geq 0~(n\in\mathcal{N})$ and $\beta\geq 0$ are
the Lagrange multipliers for the constraints
(\ref{eqn_const_target rate_2}) and (\ref{eqn_const_rho}),
respectively. If $\{\rho_n^*\}$ and $\{s_{k,n}^*\}$ are the
optimal solution of \textbf{P3}, they should satisfy the
Karush-Kuhn-Tucker (KKT) conditions \cite{BoydV04b}, which are
necessary and sufficient for the optimality. The KKT conditions
are listed as follows:
\begin{eqnarray}\label{eqn_KKT_1}
\frac{\partial J(\ldots)}{\partial s_{k,n}}\left\{
\begin{array}{ll}
=0& \mbox{if}~s_{k,n}^*>0\\
>0& \mbox{if}~s_{k,n}^*=0
\end{array}\right., ~\forall n\in\mathcal{N}, k\in\mathcal{K} \end{eqnarray}
\begin{eqnarray}\label{eqn_KKT_2}
\frac{\partial J(\ldots)}{\partial \rho_{n}}\left\{
\begin{array}{ll}
>0& \mbox{if}~\rho_{n}^*=0\\
=0& \mbox{if}~0<\rho_{n}^*<1\\
<0& \mbox{if}~\rho_{n}^*=1
\end{array}\right., \forall n\in\mathcal{N} \end{eqnarray}
\begin{equation}\label{eqn_KKT_3}
 \lambda_n\left[\rho_n^*
\sum_{k\in\mathcal{K}}\ln\left(1+\frac{g_{k,n}
s_{k,n}^*}{\rho_n^*}\right)-R\right]=0,~\forall n\in\mathcal{N}.
\end{equation}

It can be obtained from the KKT condition (\ref{eqn_KKT_1}) that
the optimal power distribution $\{p_{k,n}^*\}$ has a water-filling
structure, and is given by
\begin{equation}\label{eqn_optimal_power_allocation}
p_{k,n}^*=\frac{s_{k,n}^*}{\rho_n^*}=\left(\lambda_n-\frac{1}{g_{k,n}}\right)^+,~\forall
k\in\mathcal{K},~n\in\mathcal{N},
\end{equation}
where $(x)^+\triangleq\max(0,x)$, and $\lambda_n$ can be regarded as
the water level on hop $n$. Different hops may have different water
levels. For each hop, more power is allocated to the subcarrier with
higher channel gain and vice versa.

Let $\mathcal{K}_n$ denote the set of subcarriers over hop $n$ that
are assigned with non-zero power, i.e. satisfying $g_{k,n}>
1/\lambda_n$, $\forall k\in\mathcal{K}_n$, and let $k_n$ be the size
of the set. The subcarriers in the set are said to be active
subcarriers. Note that each water level value $\lambda_n$ cannot be
zero. Otherwise $p_{k,n}^*=0$, $\forall k,n$ and, as a result, the
constraint (\ref{eqn_const_target rate_2}) cannot be satisfied.
Hence, we obtain the closed-form expression for $\rho_n^*$ by
substituting (\ref{eqn_optimal_power_allocation}) into the KKT
condition (\ref{eqn_KKT_3}):
\begin{equation}\label{eqn_rho_lambda}
\rho_n^*\triangleq
h_n(\mathbf{g},\lambda_n)=\frac{R}{\sum_{k\in\mathcal{K}_n}\ln
g_{k,n}+k_n\ln\lambda_n},~\forall n\in\mathcal{N}.
\end{equation}
From (\ref{eqn_rho_lambda}), it can be shown that $\rho_n^*$ is
monotonically decreasing in $\lambda_n$ (note that $k_n$ also
depends on $\lambda_n$).

In the following, we derive the relation between $\lambda_n$ and
$\beta$. Taking the derivative of Lagrangian of \textbf{P3} in
(\ref{eqn_Lagrangian}) with respect to $\rho_n$, we have
\begin{equation}\label{eqn_deviation_lagrangian}
\frac{\partial J (\ldots)}{\partial
\rho_n}=\lambda_n\left[\sum_{k\in\mathcal{K}}\ln\left(1+\frac{g_{k,n}s_{k,n}}{\rho_n}\right)-\sum_{k\in\mathcal{K}}\frac{g_{k,n}
s_{k,n}}{\rho_n+g_{k,n}s_{k,n}}\right]-\beta.
\end{equation}
Suppose that there exists an $n\in\mathcal{N}$ such that
$\rho_n^*=0$ or 1, then the condition (\ref{eqn_KKT_3}) would be
violated. Thus, we have $0<\rho_n^*<1~, \forall n\in\mathcal{N}$.
Substituting (\ref{eqn_optimal_power_allocation}) into
(\ref{eqn_deviation_lagrangian}) and using (\ref{eqn_KKT_2}), we
express $\beta$ as a function of $\lambda_n$ given by
\begin{eqnarray}\label{eqn_beta_lambda}
\beta \triangleq f_n(\mathbf{g},
\lambda_n)=\lambda_n\left(\sum_{k\in\mathcal{K}_n}\ln g_{k,n}
+k_n\ln\lambda_n
-k_n\right)+\sum_{k\in\mathcal{K}_n}\frac{1}{g_{k,n}},~\forall
n\in\mathcal{N}.
\end{eqnarray}
It is seen from (\ref{eqn_beta_lambda}) that finding the optimal
water levels $\{\lambda_n\}$ at a given $\beta$ are $N$ independent
problems. It can be proven that $f_n(\mathbf{g},\lambda_n)$ is a
monotonically increasing function of $\lambda_n$ in the region
$\left[\min_k\left(\frac{1}{g_{k,n}}\right),+\infty\right]$ by
evaluating its derivative with respect to $\lambda_n$. Hence, the
inverse function, $\lambda_n = f_n^{-1}(\mathbf{g},\beta)$, exists
and is an increasing function of $\beta$. Therefore, the exact value
of $\lambda_n$ for a given $\beta$ can be obtained numerically using
binary search when the upper bound is known.

Substituting $\lambda_n=f_n^{-1}(\mathbf{g},\beta)$ into
(\ref{eqn_rho_lambda}), we can express $\rho_n^*$ as
$\rho_n^*=h_n(f_n^{-1}(\mathbf{g},\beta))$. Since $\rho_n^*$ is
decreasing in $\lambda_n$ and $\lambda_n$ is increasing in $\beta$,
we have that $\rho_n^*$ is decreasing in $\beta$. Therefore, the
optimal $\beta$ can also be obtained via binary search from the
constraint (\ref{eqn_const_rho}). Hence, the optimal solution
$\{\rho_n^*, s_{k,n}^*\}$ of \textbf{P3} and the resulting
$p_{\min}$ can be obtained through two nested binary searching
loops. The outer loop varies the Lagrange multiplier $\beta$ to meet
the time constraint. The inner loop searches the water level for
each hop at a given value of $\beta$ to satisfy
(\ref{eqn_beta_lambda}). The algorithm is outlined as follows.

\hspace{0.5cm}

\begin{tabular}{l}
\hline
\textbf{Two-nested Binary Search for minimum short-term power (TBS)}\\
\hline\\
\end{tabular}
{\small

\textbf{{Binary search for $\beta$}}
    \begin{enumerate}
    \item Find the upper bound and lower bound of $\beta$
    \begin{enumerate}
    \item For all $n\in\mathcal{N}$, let $\bar{\lambda}_n=\underline{\lambda}_n=\min_k\{1/g_{k,n}\}$
    \item Compute $\bar{\rho}_n=h_n(\mathbf{g},\bar{\lambda}_n)$ using (\ref{eqn_rho_lambda})
    \item If $\bar{\rho}_n>1/N$, update $\bar{\lambda}_n=2\bar{\lambda}_n$ and
    repeat Step 1)-b) and c), else, go to Step 1)-d)
    \item Set $\beta^{\min}=\max_{n\in\mathcal{N}}
    f_n(\mathbf{g},\underline{\lambda}_n)$, and $\beta^{\max}=\max_{n\in\mathcal{N}}
    f_n(\mathbf{g},\bar{\lambda}_n)$
    \end{enumerate}
    \item Set $high = \beta^{\max},~low = \beta^{\min}$
    \item {Let $center = (low +
    high)/2$ and \textbf{{binary search for $\lambda_n~(\forall n\in\mathcal{N})$}} at $\beta=center$}
        \begin{enumerate}
        \item {Find the upper bound and lower bound of $\lambda_n$,
                $\lambda_n^{\max}$ and $\lambda_n^{\min}$, respectively}
                \begin{enumerate}
                \item Let $\lambda_n^{\min}=\lambda_n^{\max}=\min_k
                \{1/g_{k,n}\}$
                \item Compute $\beta'=f_n(\mathbf{g},\lambda_n^{\max})$ using (\ref{eqn_beta_lambda})
                \item If $\beta'<\beta$, update
                $\lambda_n^{\min}=\lambda_n^{\max}$ and $\lambda_n^{\max}=2\lambda_n^{\max}$ and repeat Step 3)-a)-ii) and iii)\\
                else, let $high_n = \lambda_n^{\max},~low_n = \lambda_n^{\min}$, and go to Step 3)-b)
                \end{enumerate}
        \item Set $center_n = (low_n +
                high_n)/2.$ If $f_n(\mathbf{g},center_n)>\beta$, let $high_n = center_n$; otherwise, let $low_n =
    center_n$
        \item Repeat Step 3)-b) until
        $high_n-low_n<\varepsilon'$ and let $\lambda_n=center_n$
        \end{enumerate}
    \item If $\sum_{n\in\mathcal{N}} h_n(\mathbf{g},\lambda_n)>1$, let $low = center$; otherwise, let $high =
    center$
    \item Repeat Step 3) and Step 4) until $high-low<\varepsilon$
    \item Using the found $\{\lambda_n\}$ and $\beta$, obtain $\rho_n^*$ and $p_{k,n}^*$ based on (\ref{eqn_rho_lambda}) and
    (\ref{eqn_optimal_power_allocation}), respectively.
    \item Compute $p_{\min}=\sum_{n\in\mathcal{N}}
    \rho_n^*(\sum_{k\in\mathcal{K}} p_{k,n}^*)$
    \end{enumerate}
}
\begin{tabular}{l}
\hline\\
\textbf{~~~~~~~~~~~~~~~~~~~~~~~~~~~~~~~~~~~~~~~~~~~~~~~~~~~~~~~~~}\\
\end{tabular}
\vspace{-0.8cm}

 In Step 1), the boundaries of $\beta$
are determined in order to proceed with the binary search in the
outer loop. From (\ref{eqn_beta_lambda}), a common Lagrange
multiplier $\beta$ is shared by all hops and it is a monotonically
increasing function of $\lambda_n$ in the region of
$\left[\min_{k\in\mathcal{K}}\left(\frac{1}{g_{k,n}}\right),+\infty\right]$
for all $n$.
 We use
$\underline{\lambda}_n=\min_{k\in\mathcal{K}}
\left(\frac{1}{g_{k,n}}\right)$ to represent the lower bound of
$\lambda_n$, then the lower bound of $\beta$ is maximum of
$f_n(\mathbf{g},\underline{\lambda}_n)$ among $N$ hops. The same
lower bound of $\lambda_n$ will also be used in Step 3) for the
inner loop. The upper bound of $\beta$ is obtained from the fact
that there exists at least an $n^*$ such that
$\rho_{n^*}\geq\frac{1}{N}$. Correspondingly, we find a water level
$\bar{\lambda}_n=h_n^{-1}(\mathbf{g},\frac{1}{N})$ for all $n$,
where $h_n^{-1}(\mathbf{g},\cdot)$ is the inverse function of
$h_n(\mathbf{g},\cdot)$. Then for hop $n^*$, we have
${\lambda}_{n^*}\leq\bar{\lambda}_{n^*}$ since $h_n^{-1}
(\mathbf{g},\cdot)$ is a decreasing function. Therefore, due to the
monotone of $f_n(\mathbf{g},\cdot)$, the upper bound of $\beta$ can
be obtained from
\begin{equation*}
\beta=f_{n^*}(\mathbf{g},\lambda_{n^*})\leq
f_{n^*}(\mathbf{g},\bar{\lambda}_{n^*})\leq \max_{n\in\mathcal{N}}
f_n(\mathbf{g},\bar{\lambda}_n).
\end{equation*}
The algorithm then updates $\beta$ using binary chop until the sum
of the corresponding time-sharing fraction converges to 1. The
convergence of the outer loop is guaranteed by the fact that the
actual sum of time-sharing fractions is also monotonically
decreasing in $\beta$.

The aim of the inner loop in Step 3) is to find $\lambda_n~(\forall
n\in\mathcal{N})$ for a given $\beta$. We first initialize the upper
bound of $\lambda_n$,
$\lambda_n^{\max}=\min_k\left\{\frac{1}{g_{k,n}}\right\}$ and then
keep increasing it until the corresponding $\beta'$ goes beyond the
given $\beta$. In each iteration, the binary search guesses an
halfway $\lambda_n$ between the new $high$ and $low$ and repeats it
until the actual $\beta'$ approach the given $\beta$. The iteration
converges because of the monotone of $\beta$ in $\lambda_n$.

The outer loop involves $\log_2
\left(\frac{\beta^{\max}-\beta^{\min}}{\varepsilon}\right)$
iterations where $\varepsilon$ represents outer loop accuracy. The
inner loop has $N$ binary searches, and each involves $\log_2
\left(\frac{\lambda_n^{\max}-\lambda_n^{\min}}{\varepsilon'}\right)$
iterations, where $\varepsilon'$ is the inter loop accuracy. It is
observed from (\ref{eqn_rho_lambda}) and (\ref{eqn_beta_lambda})
that $\beta^{\max}=\mathcal{O}\left(NR\mbox{e}^{NR/K}\right)$ and
$\lambda_n^{\max}=\mathcal{O}\left(\mbox{e}^{NR/K}\right)$ when the
target rate is so high that all subcarriers are active. Therefore,
the average computational complexity of this algorithm is upper
bounded by the magnitude of $\frac{N^3R^2}{K^2}
\ln(\frac{NR}{\varepsilon}) \ln(\frac{1}{\varepsilon'})$ in the
asymptotic sense with a high target rate.

\subsubsection{Sub-optimal time and power
allocation}\label{subsubsec_subopt}

In the optimal time and power allocation, it is infeasible to obtain
an closed-form expression for the solution. In this part, we will
observe that when the target rate is sufficiently large, the optimal
transmission time can be approximated by an explicit function of
geometric mean of channel gains averaged on active subcarriers and
the number of active subcarriers. In addition, the product of water
level and the number of active subcarriers for each hop tends to be
the same. In the following, we shall investigate this sub-optimal
solution and show that it has a low computational complexity and
little signalling exchange.

Let $\{\rho_n,~n\in\mathcal{N}\}$ be any given time allocation that
satisfies $\sum_{n\in\mathcal{N}} \rho_n=1$ and $0\leq\rho_n\leq 1$.
The optimal power distribution at the given
 $\{\rho_n,~n\in\mathcal{N}\}$ is expressed by
(\ref{eqn_optimal_power_allocation}). Substituting
(\ref{eqn_optimal_power_allocation}) into
(\ref{eqn_real_transmission_rate}) and letting
$p_n=\sum_{k\in\mathcal{K}} p_{k,n}$, the close-form expression of
$\lambda_n$  can be obtained as \cite{TaoLZ06c}
\begin{equation}
\label{eqn_inverse_lambda}
{\lambda_n}=\left(\frac{\mbox{e}^{\frac{R}{\rho_n}}}{\prod_{k\in\mathcal{K}_n}g_{k,n}}\right)^{{1}/{k_n}},
\end{equation}
where $k_n$ is the size of the set $\mathcal{K}_n=\{k\mid g_{k,n}>
1/\lambda_n\}$. Moreover, substituting (\ref{eqn_inverse_lambda})
back into (\ref{eqn_optimal_power_allocation}), we have
\begin{equation}\label{eqn_ineqn_p_n}
p^*_n(\rho_n)=\mbox{e}^{\frac{R}{k_n}\frac{1}{\rho_n}-a_n}-b_n,
\end{equation}
where, for notation brevity, we define
\begin{equation}\label{eqn_a_n}
a_n\triangleq\frac{1}{k_n}\left(\sum_{k\in\mathcal{K}_n}\ln
g_{k,n}\right)-\ln k_n=\ln \tilde{g}_n-\ln k_n,
\end{equation}
and
\begin{equation}\label{eqn_b_n}
b_n\triangleq
\sum_{k\in\mathcal{K}_n}\frac{1}{g_{k,n}}=\frac{k_n}{\bar{g}_n}.
\end{equation}
In (\ref{eqn_a_n}) and (\ref{eqn_b_n}), $\tilde{g}_n$ and
$\bar{g}_n$ represent the geometric mean and harmonic mean of
$g_{k,n}$ over the active subcarriers at hop $n$, respectively.

For the moment, let us assume that $k_n$'s are fixed and then both
$a_n$ and $b_n$ are constants. Then, the problem of minimizing
total power for supporting the target end-to-end transmission rate
can be reformulated as \textbf{P4} only with optimization
variables $\{\rho_n,~n\in\mathcal{N}\}$
\begin{eqnarray}
\label{eqn_p_min}
\textbf{P4}:~~p_{\min}(\mathbf{g})&=&\min_{\{\rho_n\}}\sum_{n\in\mathcal{N}}\rho_n
p^*_n =\min_{\{\rho_n\}}\sum_{n\in\mathcal{N}}\rho_n\left(
\mbox{e}^{\frac{R}{k_n}\frac{1}{\rho_n}-a_n}-b_n\right)\\
\label{eqn_p_min_constraint}
 &&\mbox{s.t.} \sum_{n\in\mathcal{N}}
\rho_n =1.
\end{eqnarray}
Problem \textbf{P4} can be also solved using Lagrange multiplier
method since it is convex. The Lagrangian of this problem is given
by
\[
L(\vec{\rho},\nu)=\sum_{n\in\mathcal{N}}\left(\rho_n
\mbox{e}^{\frac{R}{k_n}\frac{1}{\rho_n}-a_n}-b_n\right)+\nu\left(1-\sum_{n\in\mathcal{N}}\rho_n\right),
\]
where the lagrange multiplier $\nu$ satisfies constraint
(\ref{eqn_p_min_constraint}). Applying KKT condition, the optimal
time-sharing fraction $\rho_n$ should satisfy
\begin{equation}\label{eqn_dev_lagrange}
\frac{\partial L(\vec{\rho},\nu)}{\partial
\rho_n}=\mbox{e}^{\frac{R}{k_n}\frac{1}{\rho_n}-a_n}-\frac{R}{k_n}\frac{1}{\rho_n}\mbox{e}^{\frac{R}{k_n}\frac{1}{\rho_n}-a_n}-\nu=0.
\end{equation}
 The closed-form solution to
(\ref{eqn_dev_lagrange}) is difficult to obtain.

It is known that when the target transmission rate is sufficiently
small, the power saving through time adaptation is insignificant
\cite{YaoG05j}. This result motivates us to focus on the high
target transmission rate with $R\gg K$. We consider two particular
hops, $n_1$ and $n_2$. Under the assumption of a high target
transmission rate, the equation (\ref{eqn_dev_lagrange}) can be
approximated by
\[
\mbox{e}^{-a_{n_1}}\frac{R}{k_{n_1}\rho_{n_1}}
\mbox{e}^{\frac{R}{k_{n_1}\rho_{n_1}}}\approx
\mbox{e}^{-a_{n_2}}\frac{R}{k_{n_2}\rho_{n_2}}
\mbox{e}^{\frac{R}{k_{n_2}\rho_{n_2}}}.
\]
From the above approximation, we can obtain a ratio
\begin{equation}\label{eqn_ratio}
\frac{k_{n_1}\rho_{n_1}}{k_{n_2}\rho_{n_2}}\approx
1+\frac{k_{n_1}\rho_{n_1}}{R}(a_{n_2}-a_{n_1})-\frac{k_{n_1}\rho_{n_1}}{R}\ln\left(\frac{k_{n_1}\rho_{n_1}}{k_{n_2}\rho_{n_2}}
\right).
\end{equation}
Without loss of generality, we assume $a_{n_2}\geq a_{n_1}$, then
we have $(k_{n_1}\rho_{n_1})/(k_{n_2}\rho_{n_2})\geq 1$ from
(\ref{eqn_ratio}). Thus, (\ref{eqn_ratio}) leads to
\begin{equation}\label{eqn_ineqn1}
1\leq\frac{k_{n_1}\rho_{n_1}}{k_{n_2}\rho_{n_2}}\leq
1+\frac{k_{n_1}\rho_{n_1}}{R}(a_{n_2}-a_{n_1}).
\end{equation}
Using the inequality $\ln(x)\leq x-1$ and (\ref{eqn_ratio}), we
obtain a lower bound of $(k_{n_1}\rho_{n_1})/(k_{n_2}\rho_{n_2})$
after some manipulations,
\begin{equation}\label{eqn_ineqn2}
\frac{k_{n_1}\rho_{n_1}}{k_{n_2}\rho_{n_2}}\geq
1+\frac{k_{n_1}\rho_{n_1}(a_{n_2}-a_{n_1})}{R\left(1+\frac{k_{n_1}\rho_{n_1}}{R}\right)}.
\end{equation}
Since $R\gg K$, inequalities (\ref{eqn_ineqn1}) and
(\ref{eqn_ineqn2}) lead to
\[
\frac{k_{n_1}\rho_{n_1}}{k_{n_2}\rho_{n_2}}\approx
1+\frac{k_{n_1}\rho_{n_1}}{R}(a_{n_2}-a_{n_1}).
\]
After manipulation, we have
\[
\frac{1}{k_{n_2}\rho_{n_2}}-\frac{a_{n_2}}{R}\approx
\frac{1}{k_{n_1}\rho_{n_1}}-\frac{a_{n_1}}{R},~\forall
n_1,n_2\in\mathcal{N}.
\]
Let
\[
\frac{1}{k_{n}\rho_{n}}-\frac{a_{n}}{R}=\mu,~\forall
n\in\mathcal{N}.
\]
We can obtain the approximated but close-form solution to
(\ref{eqn_dev_lagrange}) as follows
\begin{equation}\label{eqn_app_rho}
\rho_n'=\frac{R}{k_n R \mu +k_n a_n}, \forall n\in\mathcal{N},
\end{equation}
where $\mu$ is determined by the time constraint
(\ref{eqn_p_min_constraint}), and can be obtained through binary
search.  Substituting (\ref{eqn_app_rho}) into
(\ref{eqn_ineqn_p_n}), the corresponding transmission power
allocated to hop $n$ is given by
\begin{equation}\label{eqn_app_p_n}
p_n'=\mbox{e}^{R\mu}-b_n.
\end{equation}
Furthermore, substituting (\ref{eqn_app_rho}) into
(\ref{eqn_inverse_lambda}) yields the sub-optimal water level for
hop $n$ as
\begin{equation}\label{eqn_sub_water_level}
\lambda_n=\frac{\mbox{e}^{R\mu}}{k_n}.
\end{equation}
 From
(\ref{eqn_optimal_power_allocation}), (\ref{eqn_app_p_n}) and
(\ref{eqn_sub_water_level}), we can regard the sub-optimal power
allocation algorithm as a two-level water filling algorithm. First,
the power is poured among all the hop according to
(\ref{eqn_app_p_n}) using the water level $\mbox{e}^{R\mu}$ and the
hop with small $b_n$ will be given more power. In each hop, the
power obtained from the previous level is then poured among
different subcarriers following
(\ref{eqn_optimal_power_allocation}), and the water level is equal
to $\mbox{e}^{R\mu}/k_n$.

Consider a special case where $\mbox{e}^{R\mu}$ is sufficiently
large so that all subcarriers are active, i.e., $k_n=K$. It follows
immediately from (\ref{eqn_app_rho}) that the hops with low
geometric mean of channel gains over the subcarriers should be
assigned with longer transmission time. Also, it follows from
(\ref{eqn_app_p_n}) one should lower the transmission power for the
hops with low harmonic mean of channel gains. An intuitive
understanding of this result is that a high priority is given the
hop with poor channel condition to take advantage of ``Lazy
Scheduling'' \cite{GamalNPUZ02c} to avoid this hop becoming the
bottleneck of the whole link. The idea behind ``Lazy Scheduling'' is
that energy required to transmit a certain amount of information
decrease when prolonging transmission time.

We now relax the assumption made earlier that $k_n$'s fixed and
propose an iterative procedure to find the best $k_n$'s for this
sub-optimal problem.\\

\begin{tabular}{l}
\hline
\textbf{Iterative Algorithm of Sub-optimal power and time allocation (IAS)}\\
\hline\\
\end{tabular}

{\small
\begin{enumerate}
\item Initialization of $k_n$ \\
Set $k_n=K,~\forall n\in\mathcal{N}$
\item Binary search for $\mu$
for a given $\{k_n,~\forall n\}$
    \begin{enumerate}
    \item Set $high=\mu_{\max}$, $low=\mu_{\min}$
    \item Let $center=(low+high)/2$, and calculate $\{\rho_n',~\forall
    n\}$ when $\mu=center$
    according to (\ref{eqn_app_rho})
    \item If $\sum_{n\in\mathcal{N}}\rho_n'>1$, let $low=center$;
    otherwise, let $high=center$
    \item Repeat Step 2)-b) and c) until
    $high-low<\varepsilon''$
    \end{enumerate}
\item Find $k_n~(\forall n)$ in the set $\{1,\ldots,K\}$ for a
given $\rho_n'$ to meet the target rate $R$ based on (\ref{eqn_r_d_rho})%
\item Repeat Step 2) and 3) until $k_n$'s are unchanged %
\item Compute $p'_{k,n}$ through substituting
(\ref{eqn_inverse_lambda}) into
(\ref{eqn_optimal_power_allocation}) \item Obtain the required
total power $p_{\min}'=\sum_{n}\rho_n'(\sum_k p_{k,n}')$
\end{enumerate}}
\begin{tabular}{l}
\hline\\
\textbf{~~~~~~~~~~~~~~~~~~~~~~~~~~~~~~~~~~~~~~~~~~~~~~~~~~~~~~~~~~~~~~~~~~~}\\
\end{tabular}
\vspace{-0.8cm}

In Step 2)-a) $\mu_{\max}$ and $\mu_{\min}$ represent the upper
bound and lower bound of $\mu$, respectively. The exact value
$\mu_{\min}=\max_{n\in\mathcal{N}}
\left(\frac{1}{k_n}-\frac{a_n}{R}\right)$ can be obtain from the
time constraint $0\leq \rho_n'\leq 1$. Its upper bound
$\mu_{\max}$ could be
$\min_{n\in\mathcal{N}}\left(\frac{NR-k_na_n}{k_nR}\right)$, since
if
$\mu>\min_{n\in\mathcal{N}}\left(\frac{NR-k_na_n}{k_nR}\right)$,
$\rho_n'<1/N~(\forall n)$ from (\ref{eqn_app_rho}), which violates
the constraint $\sum_n \rho_n'=1$.

The implementation of  the above algorithm can be done as follows.
At the beginning of each time frame, we first assume that the
transmission is on for all subcarriers. The central controller
searches for $\mu$ and broadcast it to all relays. The relays and
source node then compute their own transmission time
$\{\rho_n',~\forall n\in\mathcal{N}\}$ using (\ref{eqn_app_rho})
locally. The required power allocation for hop $n$ to meet the
target rate should satisfy
\begin{equation}\label{eqn_r_d_rho}
\sum_{k\in\mathcal{K}} \ln(1+g_{k,n}p_{k,n})
=\sum_{k\in\mathcal{K}_n}
\ln(g_{k,n}\lambda_n)\stackrel{(a)}{=}\frac{R}{\rho_n'},~\forall
n\in\mathcal{N}.
\end{equation}
The left side of the above equation (a) can be shown to be a
monotonically increasing function of $\lambda_n$, and is denoted as
$z_n(\lambda_n)$. With loss of generality, we assume $g_{1,n}\geq
g_{2,n} \geq \ldots \geq g_{K,n}~(\forall n\in\mathcal{K})$. Each
$\lambda_n$ maps to a unique $k_n$ which satisfies that $
\frac{1}{g_{k_n,n}}\leq \lambda_n \leq \frac{1}{g_{k_n+1,n}}$. Thus,
we have
\[
z_n\left(\frac{1}{g_{k_n,n}}\right)\leq \frac{R}{\rho_n'} \leq
z_n\left(\frac{1}{g_{k_n+1,n}}\right).
\]
Therefore, the desired $k_n$ in Step 3) can be obtain through
binary search in the set of $\{1,\ldots, K\}$ by comparing
$z_n(1/g_{k_n,n})$ with $R/\rho_n'$.
The found $k_n$ and the geometric and harmonic mean of channel
gains on these $k_n$ subcarriers are returned to the input of
(\ref{eqn_app_rho}) in the central controller. This procedure
repeats until the $k_n$'s are unchanged. Although the convergence
of this algorithm cannot be guaranteed theoretically, divergent
behaviors were never observed in the simulation. In the following,
we shall use simulation to examine the average number of
iterations for the algorithm to converge and the average required
short-term total transmission power.

In the simulation, SUI channel model for the fixed broadband
wireless access channel environments \cite{ErcegH01s} is used and
the channel parameters will be detailed in Section
\ref{sec_numerical}. The simulation is run for $10^3$ time frames
to evaluate the average performance. The number of subcarriers is
set to 16.

\begin{figure}
\centering
\includegraphics[width=5in]{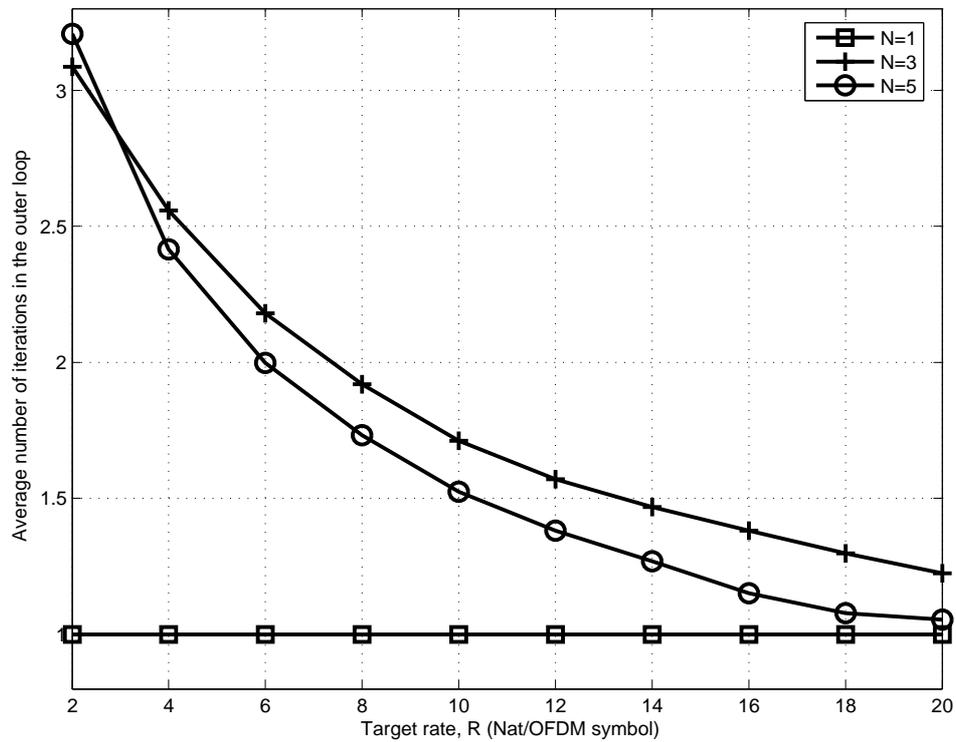}
\caption{Average number of iterations in the outer loop required
for the search of $\{k_n\}$} \label{fig_APTA_app_ave_M}
\end{figure}

Fig. \ref{fig_APTA_app_ave_M} shows the average iterations in the
outer loop over $10^3$ independent channel realizations required
for the search of $\{k_n\}$ to converge. It is shown that the
average iteration numbers, denoted as $M$, is decreasing in $R$
and approaches $1$ when the target rate is sufficiently large. It
can be explained by the fact that $k_n=K$ when $R$ goes infinity.

\begin{figure}
\centering
\includegraphics[width=5in]{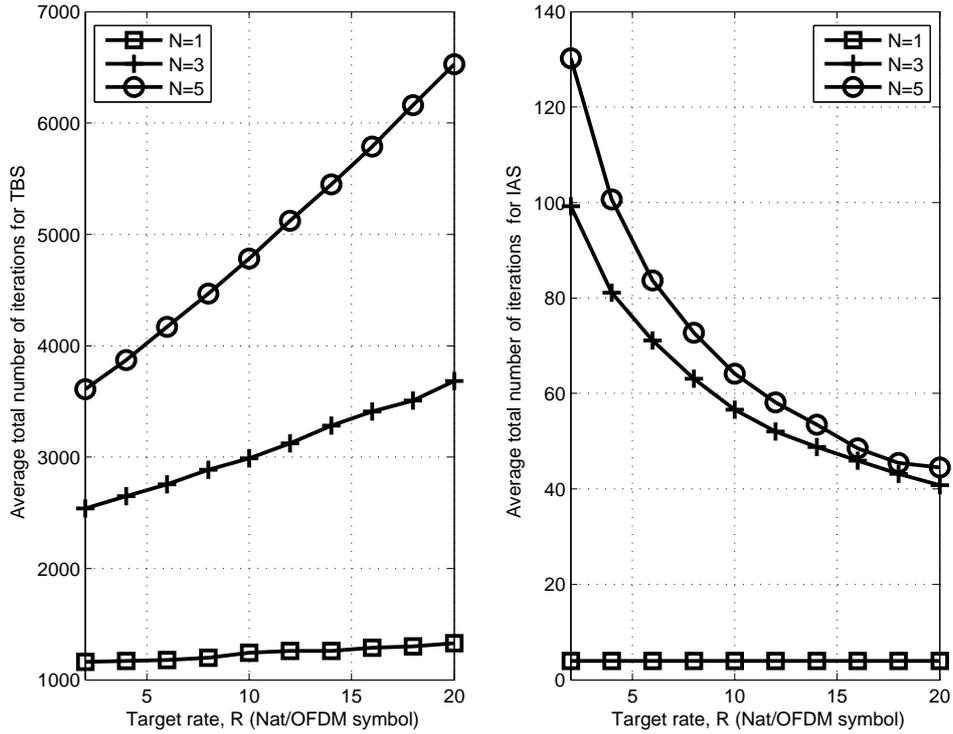}
\caption{Average total number of iterations using TBS and IAS}
\label{fig_total_complexity}
\end{figure}

Since the binary search for $\mu$ in the inner loop involves
$\log_2\left(\frac{\mu_{\max}-\mu_{\min}}{\varepsilon''}\right)$
iterations and finding $k_n$ for a given $R/\rho_n$ involves
$\log_2(K)$ ones, the total number of iterations required for the
IAS can be express as
\begin{equation}
C_{IAS}=M\left\{\log_2\left[\frac{\min_{n\in\mathcal{N}}\left(\frac{NR-k_na_n}{k_nR}\right)-\max_{n\in\mathcal{N}}
\left(\frac{1}{k_n}-\frac{a_n}{R}\right)}{\varepsilon''}\right]+N\log_2(K)\right\}
\end{equation}
Since $M$ is decreasing in $R$, $C_{IAS}$ is also decreasing in
$R$ and upper bounded by a linear function of $N$. Fig.
\ref{fig_total_complexity} compares average total complexities
between TBS and IAS for different $R$ and different $N$.

\begin{figure}
\centering
\includegraphics[width=5in]{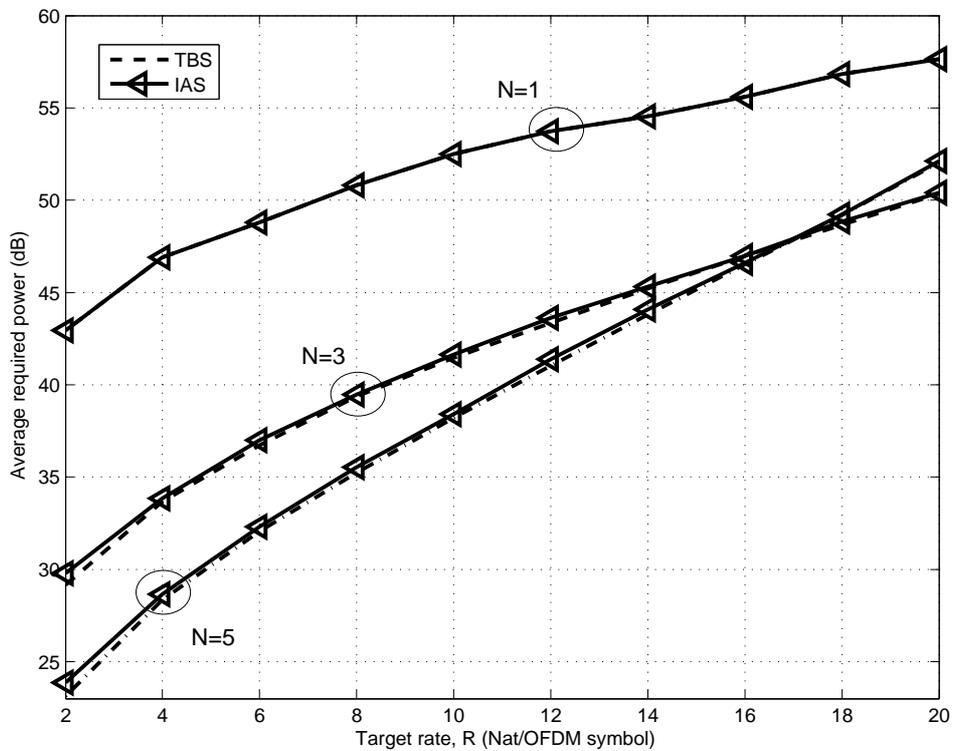}
\caption{Average short-term power required to meet the target
rate, $R$} \label{fig_APTA_app_ave_power}
\end{figure}

Fig. \ref{fig_APTA_app_ave_power} compares the average power
required to meet the target rate between TBS developed in Section
\ref{subsubsec_opt} and its sub-optimal algorithm, IAS. It is
shown that IAS serves as a good approximation of TBS, especially
for a high target rate.

As we discuss previously, the required controlling signals from
the feedback channel are only geometric mean, harmonic mean of
$g_{k,n}$ and the number of active subcarriers over each hop
instead of $\{g_{k,n}~,\forall k\in\mathcal{K}, n\in\mathcal{N}\}$
as in TBS, thus the signalling exchange is greatly reduced when
the number of subcarriers is large and/or the target rate is high.
Furthermore, since IAS has low complexity and near-optimal power
consumption performance, it is a good candidate for a sufficiently
high target rate in a real system.

\subsection{Long-Term Power Threshold Determination}\label{subsec_B}
We have discussed the short-term total transmission power
minimization. If the transmission is on for every possible channel
realization, the long term power constraint may be violated.
Similar to the single user case \cite{CaireTB99j}, the optimal
power allocated to all hops for \textbf{P1} with a long term power
constraint must have the following structure,
\begin{eqnarray}\label{eqn_power_structure}
p(\mathbf{g})=\left\{ \begin{array}{ll}
p_{\min}(\mathbf{g})& \mbox{with probability}~w(\mathbf{g})\\
 0 &\mbox{with probability}~1-w(\mathbf{g})
\end{array}\right.. \end{eqnarray}
Thus, the outage probability is $P(r(\mathbf{g}, \vec{\rho},
\mathbf{p}) < R) = \mathbb{E}[1- w(\mathbf{g})]$. Then solving
\textbf{P1} is equivalent to finding the optimal weighting function
$w(\mathbf{g})$ to the following problem,
\begin{eqnarray}
\min_{w(\mathbf{g})}&&\mathbb{E}[1- w(\mathbf{g})]\nonumber\\
\mbox{s.t.} && 0\leq w(\mathbf{g})\leq 1\nonumber\\
&& \mathbb{E}[p_{\min}(\mathbf{g})w(\mathbf{g})]=P.\nonumber
\end{eqnarray}
According to the result of \cite[Lemma 3]{CaireTB99j}, the optimal
weighting function has the form
\begin{eqnarray}\label{eqn_w}
w^*(\mathbf{g})=\left\{ \begin{array}{ll}
1& \mbox{for}~p_{\min}(\mathbf{g})<s^*\\
w_0 &\mbox{for}~p_{\min}(\mathbf{g})=s^*\\
0&\mbox{for}~p_{\min}(\mathbf{g})>s^*
\end{array}\right.. \end{eqnarray}
The power threshold $s^*$ is given by
$
s^*=\sup \{s:\mathcal{P}(s)<P\},
$
and $w_0$ is given by
$
w_0=\frac{P-\mathcal{P}(s^*)}{\mathcal{\bar{P}}(s^*)-\mathcal{P}(s^*)},
$
where the region $\mathcal{R}(s)$ and $\bar{\mathcal{R}}(s)$ are
defined as
$
\mathcal{R}(s)=\{\mathbf{g}:p_{\min}(\mathbf{g})<s\},~\bar{\mathcal{R}}(s)=\{\mathbf{g}:p_{\min}(\mathbf{g})\leq
s\},
$
and the corresponding average power over the two sets are:
$
\mathcal{P}(s)=\mathbb{E}_{\mathbf{g}\in\mathcal{R}(s)}[p_{\min}(\mathbf{g})],
\bar{\mathcal{P}}(s)=\mathbb{E}_{\mathbf{g}\in\bar{\mathcal{R}}(s)}[p_{\min}(\mathbf{g})]
$
The resulting minimum outage probability is denoted as
\[
P_{\rm{APT}}^{{out}} =
1-\mbox{Prob}\{\mathbf{g}\in\mathcal{R}(s^*)\}- w_0
\mbox{Prob}\{p_{\min}(\mathbf{g}) = s\}.
\]
From (\ref{eqn_power_structure}) and (\ref{eqn_w}), we see that
when the minimum total power for all hops required to support the
target transmission rate is beyond the threshold $s^*$,
transmission is turned off. When the required power is less than
the threshold, the transmission follows the minimum transmission
power strategy derived from Section \ref{subsec_A}.

The value of $s^*$ can be computed a priori if the fading
statistics are known. Otherwise, the threshold can be estimated
using fading samples. During the estimation of the threshold,
since the channel is assumed to be ergodic, the ensemble average
transmission power is equal to the time average
\[
\mathbb{E}_{\mathbf{g}\in\mathcal{R}(s)}[p_{\min}(\mathbf{g})]=\lim_{t\rightarrow\infty}\frac{1}{t}\sum_{i=1}^t
\hat{p}(i),
\]
where $\hat{p}_{(i)}$ represents the actual transmission power at
time frame $i$. Thus, the threshold is always adjusted in the
opposite direction of $P - \frac{1}{t}\sum_{i=1}^{t}\hat{p}_{(i)}$
as
\begin{equation}\label{eqn_update_s}
s^*(t+1)=s^*(t)\left[1+\epsilon\left(P-\frac{1}{t}\sum_{i=1}^{t}\hat{p}_{(i)}\right)\right].
\end{equation}
where $t$ is the time frame index.

Combining the short-term power minimization and long-term power
threshold determination, the full algorithm for APT is outlined as
follows.\\

\hspace{0.5cm} {\small
\begin{tabular}{l}
\hline
\textbf{APT}~~\\
\hline\\
\end{tabular}
\begin{enumerate}
\item Set $t = 1$  and $s_{(t)}^* = P$%
\item Search for minimum short-term power
(developed in Section \ref{subsec_A})%
\item \emph{On-off decision}\\
If $p_{\min}> s^*_{(t)}$, turn off the transmission and let
$\hat{p}_{(t)} = 0$; otherwise, turn on the transmission and let
$\hat{p}_{(t)} = p_{\min}$.%
\item \emph{Update the threshold $s^*$}
\begin{equation}\label{eqn_update_s}
s^*(t+1)=s^*(t)\left[1+\epsilon\left(P-\frac{1}{t}\sum_{i=1}^{t}\hat{p}_{(i)}\right)\right].
\end{equation}
\item Let $t= t + 1$ and return to Step 1).
\end{enumerate}
\begin{tabular}{l}
\hline\\
\textbf{~~~~~~~~~~~~~~~~~~~~~~~~~~~~~~~~~~~~~~~~~~~~~~~~~~~~~~~~~~~~~~~~~~~}\\
\end{tabular}
\vspace{-0.8cm} }

 If TBS in Section \ref{subsubsec_opt} is used in Step 2), we name
 the optimal APT as APT-opt for short. If IAS in Section
 \ref{subsubsec_subopt} is used, we denote it as APT-sub.

\subsection{Special case: adaptive power and fixed time allocation}

APFT can be viewed as a special case of APT by fixing $\rho_n =
1/N$. It can also be solved following two steps: short-term power
minimization and long-term power threshold determination. However,
unlike APT, the first step can be performed locally, i.e., each
transmitter only needs to know the local CSI over the associated hop
to solve the problem
\begin{eqnarray}
\min_{\{p_{k,n}\}}&& \sum_{k\in\mathcal{K}} p_{k,n}\nonumber\\
\mbox{s.t.}&& \frac{1}{N} \sum_{k\in\mathcal{K}} \ln(1+g_{k,n}
p_{k,n})\geq R,\nonumber
\end{eqnarray}
for all $n\in\mathcal{N}$. The solution of the problem is easily
obtained as (\ref{eqn_optimal_power_allocation}), where the water
level is given by (\ref{eqn_inverse_lambda}) with $\rho_n = 1/N$.

\section{Numerical results}\label{sec_numerical}
In this section, we present some numerical results to illustrate the
performance of the proposed adaptive power and time allocation for
OFDM based linear relay networks. The proposed algorithms, APT-opt
and APT-sub, are compared with UPT, FPAT and APFT as defined in
Section \ref{sec_system_model}.

\begin{figure}
\centering
\includegraphics[width=5in]{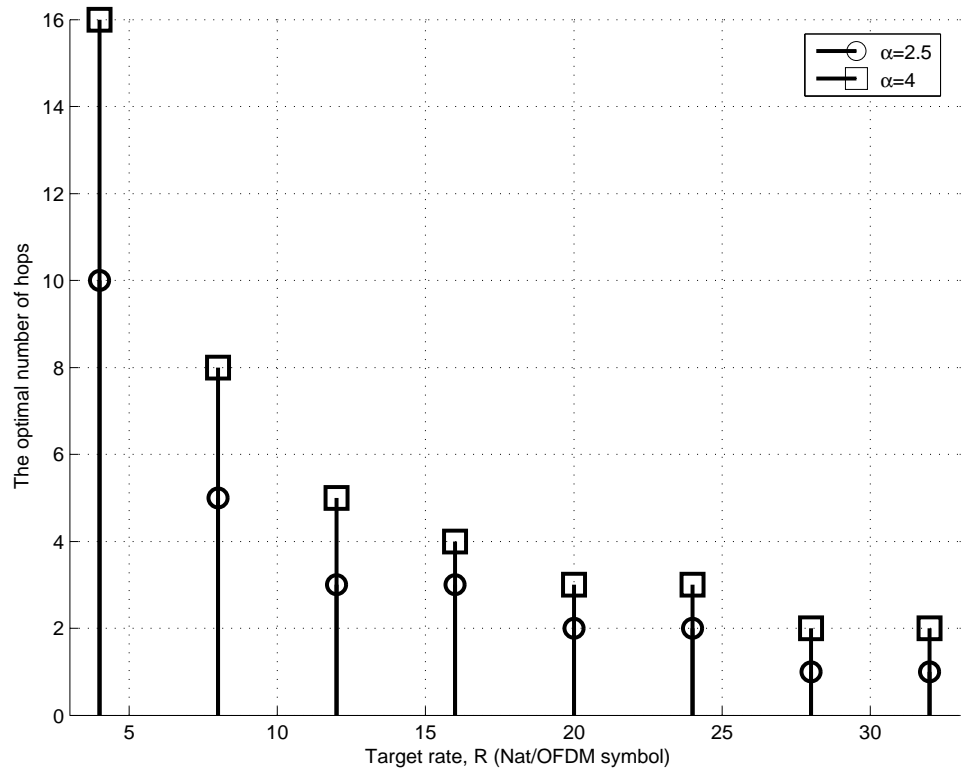}
\caption{  The optimal number of hops vs. target rate
 under APTA when $\alpha=2.5$ and 4}
\label{fig_opt_num_hop}
\end{figure}

We consider an $N$-hop linear wireless network. The acceptable BER
is chosen to be $10^{-5}$, which corresponds to 8.2dB SNR gap. We
fix the bandwidth to be 1MHz and the end-to-end distance to be
1km. The relays are equally spaced. In all simulations, the
channel over each hop is modelled by Stanford University Interim
(SUI)-3 channel model with a central frequency at around 1.9 GHz
to simulate the fixed broadband wireless access channel
environments \cite{ErcegH01s}. The SUI-3 channel is a 3-tap
channel. The received signal fading on the first tap is
characterized by a Ricean distribution with K-factor equal to 1.
The fading on the other two taps follows a Rayleigh distribution.
The root-mean-square (rms) delay spread is 0.305$\mu$s. Then the
coherence bandwidth is approximately 65KHz. Hence,
 the number of subcarrier $K$ should be
greater than 15.2 so that the subcarrier bandwidth is small enough
to experience the flat fading. Here we choose $K=16$. Doppler
maximal frequency is set to 0.4 Hz. Intermediate path loss
condition (\cite[Category B]{ErcegGTP99j}) is chosen as the path
loss model, which is given by $PL=A+\alpha
\lg\left(\frac{d}{d_n}\right),$ where $A = 20 \lg(4\pi
d_0/\lambda)$ ($\lambda$ being the wavelength in m), $\alpha$ is
the path-loss exponent with $\alpha = (a-b h_b + c / h_b)$. Here
$h_b=30m$ is chosen as the height of the base station , $d_0 =
100m$ and $a$, $b$, $c$ are 4, 0.0065 and 17.1
 given in \cite{ErcegGTP99j}. The corresponding $\alpha$ will be
 used in all simulations except the one in Fig. \ref{fig_opt_num_hop}. In each simulation, $10^4$ time frames are used to estimate the
outage probability.

\begin{figure}
\centering
\includegraphics[width=5in]{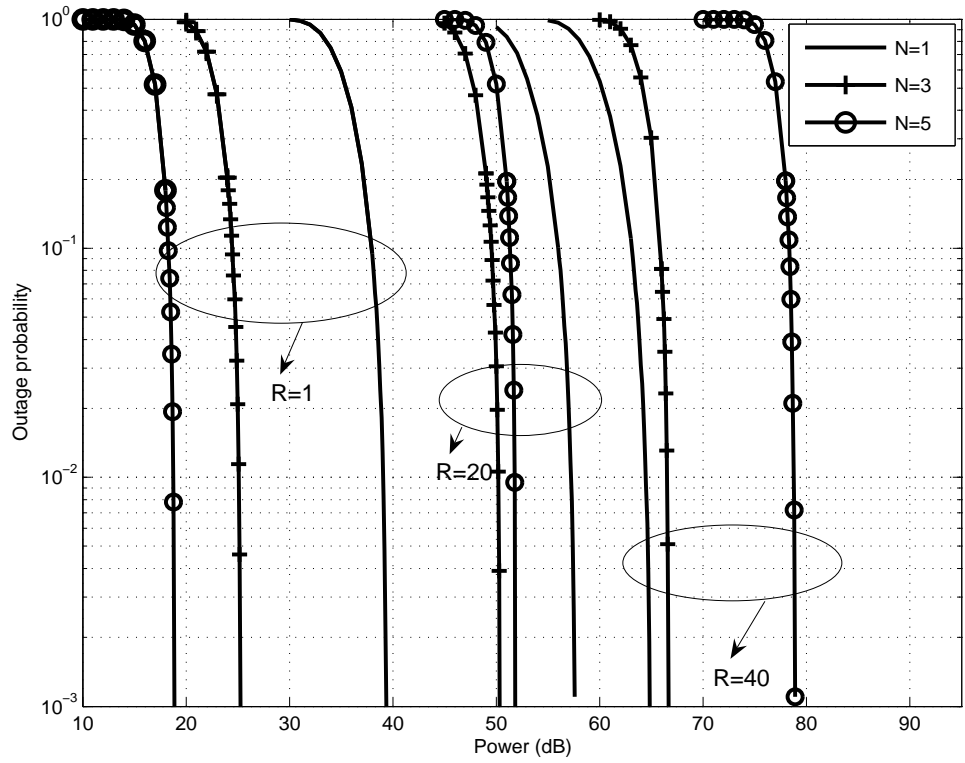}
\caption{  End-to-end outage probability vs. average total
transmission power
 under APTA when $K=16$}
\label{fig_APTA_outage_pro}
\end{figure}

Fig. \ref{fig_APTA_outage_pro} shows the end-to-end outage
probabilities versus average total transmission power for $R =1, 20$
and 40 Nat/OFDM symbol using APT-opt when $N$ varies in the set of
$\{1, 3, 5\}$. From the figure, it is shown that multi-hop
transmission can help to save total power consumption when the
target transmission rate is low (e.g., $R = 1$) whereas it is better
to send data directly to the destination if the target transmission
rate is high (e.g., $R = 40$). That can be explained by the
following fact. As the number of hops increases, the path loss
attenuation on each hop reduces. But the transmission time spent at
each hop also reduces since the total frame length is fixed. It is
observed from (\ref{eqn_real_transmission_rate}) that the
transmission rate is linear in transmission time and concave in
channel gain. Hence, when the target transmission rate increases,
the loss due to transmission time reduction cannot be evened out by
the gain brought by path loss reduction.

Fig. \ref{fig_opt_num_hop} shows the optimal number of hops to
achieve minimum power consumption at different target transmission
rates. Here, the outage probability is fixed to 1$\%$, and the path
loss exponent $\alpha=$2.5 and 4, respectively. It is observed that
the optimal number of hops is roughly proportional to the inverse of
$R$, and increasing linearly in $\alpha$. A similar trend is shown
in \cite{SikoraLHCF04c} where a spacial case, frequency-flat fading
channel and a fixed short-term total power constraint, is
considered.

\begin{figure}
\centering
\includegraphics[width=5in]{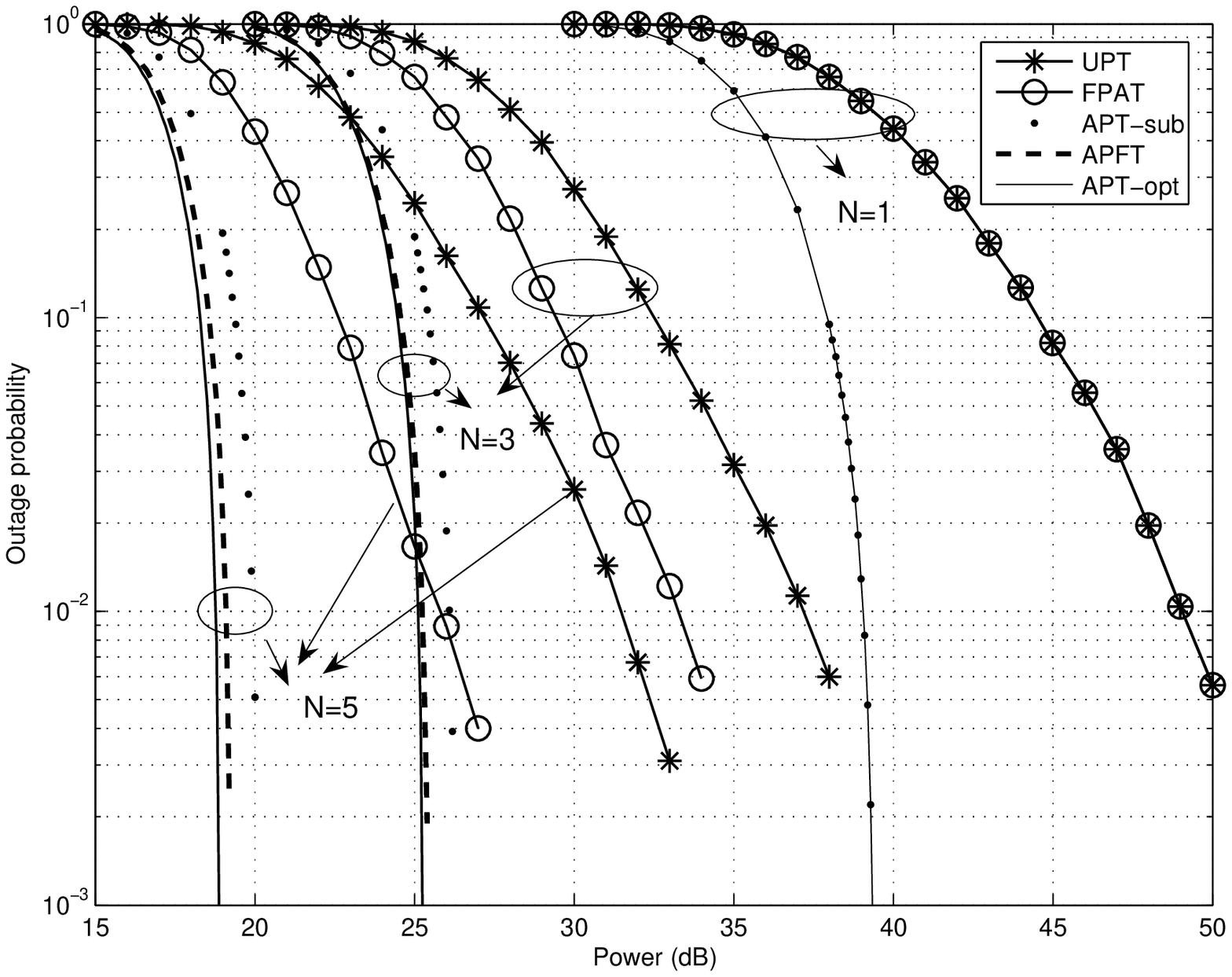}
\caption{End-to-end outage probability vs. average total
transmission power under APT-opt, APT-sub, APFT, FPAT and UPT when
$K=16$ and $R=1$ Nat/OFDM symbol} \label{fig_outage_compare_R1}
\end{figure}

\begin{figure}
\centering
\includegraphics[width=5in]{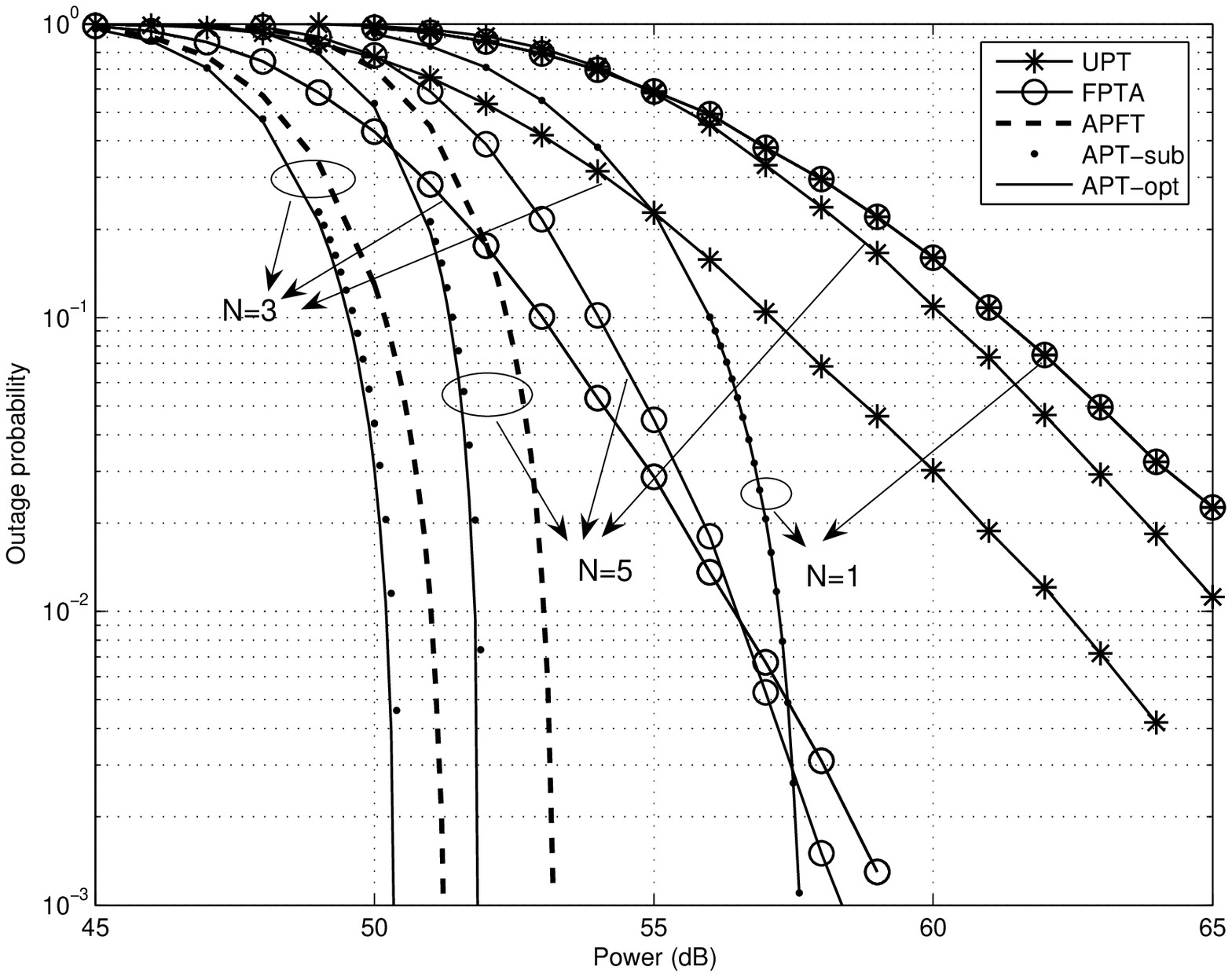}
\caption{End-to-end outage probability vs. average total
transmission power under APT-opt, APT-sub, APFT, FPAT and UPT when
$K=16$ and $R=20$ Nat/OFDM symbol} \label{fig_outage_compare_R20}
\end{figure}

Fig. \ref{fig_outage_compare_R1} and Fig.
\ref{fig_outage_compare_R20} compare the end-to-end outage
probabilities achieved by different power and time adaptation
schemes for $R=1$ and $20$ Nat/OFDM symbols, respectively. A number
of interesting observations can be made from the two figures.
First, by comparing the curves of FPAT and UPT it is observed that
just adapting per-hop transmission time alone can increase the
performance considerably. But the decreasing speed of the outage
probability as the total power increases is not increased much.
On the other hand, by comparing the curves of APFT and UPT, it is
seen that power adaption can bring dramatic improvement on the
performance. In particular, the slope of the outage probability
curves approaches almost infinity. This indicates that by turning
off the transmission when the channel suffers from deep fade can
achieve significant power saving.
Next, comparing APT-opt with APFT we can see that time adaptation on
top of power adaptation is still beneficial, but the gain is rather
limited when the target data rate is small.
Finally, it can be seen that the performance of APT-sub is even
worse than that of APFT when the target rate is low (e.g. $R=1$).
But for large target rate ($R=20$), APT-sub becomes superior and is
near optimal.

%
%
%
%

The above numerical results suggest that multi-hop transmission is
favorable at low and medium target rates, whereas a direct
transmission from source to destination is preferred if the target
rate is high.
Also, power adaptation plays a more important rule than time
adaptation in minimizing the end-to-end outage probability. In
particular,
APFT is a good choice in practice for low target rates since it has
similar performance with APT-opt and yet is much less complex. For
the similar reason, APT-sub is recommended at medium target rates.

\section{Conclusions}\label{sec_conclusions}
In this work, we consider adaptive power and time allocations for
OFDM based linear relay networks for end-to-end outage probability
minimization. The problem is solved in two steps. First, we derive
the minimum short-term total power to meet the target transmission
rate.
Both optimal and sub-optimal algorithms are proposed. In particular,
the sub-optimal algorithm suggests prolonging the transmission time
for the hop with low geometric mean of channel gains averaged over
subcarriers while lowering the transmission power for the hop with
low harmonic mean.
In the second step, the transmission on-off is determined by
comparing the required minimum total power with a threshold, which
is selected to satisfy the long-term total power constraint.
%
%
%
%
%
Numerical study is carried out to illustrate the performance of
different resource adaptation schemes: APT-opt, APT-sub, APFT, FPAT
and UPT. We find that the three schemes with adaptive power control,
APT-opt, APT-sub and APFT, provide significant power savings at a
same end-to-end outage probability over the other two. While APFT is
a good choice for practical implementation at low target rates,
APT-sub becomes near optimal at medium target rats.

%


\bibliographystyle{IEEEtran}
\bibliography{bibliography}

\begin{thebibliography}{10}
\providecommand{\url}[1]{#1}
\csname url@rmstyle\endcsname
\providecommand{\newblock}{\relax}
\providecommand{\bibinfo}[2]{#2}
\providecommand\BIBentrySTDinterwordspacing{\spaceskip=0pt\relax}
\providecommand\BIBentryALTinterwordstretchfactor{4}
\providecommand\BIBentryALTinterwordspacing{\spaceskip=\fontdimen2\font plus
\BIBentryALTinterwordstretchfactor\fontdimen3\font minus
  \fontdimen4\font\relax}
\providecommand\BIBforeignlanguage[2]{{%
\expandafter\ifx\csname l@#1\endcsname\relax
\typeout{** WARNING: IEEEtran.bst: No hyphenation pattern has been}%
\typeout{** loaded for the language `#1'. Using the pattern for}%
\typeout{** the default language instead.}%
\else
\language=\csname l@#1\endcsname
\fi
#2}}

\bibitem{PabstWSH04j}
R.~Pabst, B.~Walke, D.~Schultz, P.~Herhold, H.~Yanikomeroglu, S.~Mukherjee,
  H.~Viswanathan, M.~Lott, W.~Zirwas, M.~Dohler, H.~Aghvami, D.~Falconer, and
  G.~Fettweis, ``Relay-based deployment concepts for wireless and mobile
  broadband radio,'' \emph{{IEEE} Commun. Mag.}, vol.~42, no.~9, pp. 80--89,
  2004.

\bibitem{WiMax16j}
\BIBentryALTinterwordspacing
``Amendment to {IEEE} standard for local and metropolitan area networks - part
  16: Air interface for fixed and mobile broadband wireless access systems -
  multihop relay specification,'' March 2006. [Online]. Available:
  \url{ieee802.org/16/relay/}
\BIBentrySTDinterwordspacing

\bibitem{YaoCG05j}
Y.~Yao, X.~Cai, and G.~Giannakis, ``On energy efficiency and optimum resource
  allocation of relay transmissions in the low-power regime,'' \emph{{IEEE}
  Trans. Wirel. Commun.}, vol.~4, no.~6, pp. 2917 -- 2927, Nov. 2005.

\bibitem{SikoraLHC06j}
M.~Sikora, J.~N. Laneman, M.~Haenggi, J.~Daniel J.~Costello, and T.~E. Fuja,
  ``Bandwidth-and power-efficient routing in linear wireless networks,''
  \emph{{IEEE} Trans. Info. Theory}, vol.~52, pp. 2624--1633, 2006.

\bibitem{OymanLS06j}
O.~Oyman, J.~N. Laneman, and S.~Sandhu, ``Multihop relaying for broadband
  wireless mesh networks: From theory to practice,'' \emph{IEEE Commun. Mag.},
  vol.~45, no.~11, pp. 116--122, Nov. 2007.

\bibitem{RadunovicB03c}
B.~Radunovic and J.-Y.~L. Boudec, ``Joint scheduling, power control and routing
  in symmetric, one-dimensional, multi-hop wireless networks,'' in
  \emph{WiOpt}, France, March 2003.

\bibitem{LiL06j}
G.~Li and H.~Liu, ``Resource allocation for {OFDMA} relay networks with
  fairness constraints,'' \emph{IEEE J. Sel. Areas Commun.}, vol.~11, pp.
  2061--2069, Nov. 2006.

\bibitem{DaiGC07}
L.~Dai, B.~Gui, and L.~J. Cimini~Jr., ``Selective relaying in {OFDM} multihop
  cooperative networks,'' in \emph{Proc. {IEEE} {WCNC}}, Hong Kong, March 2007,
  pp. 963--968.

\bibitem{ZhangJT08c}
X.~Zhang, W.~Jiao, and M.~Tao, ``End-to-end resource allocation in {OFDM} based
  linear multi-hop networks,'' in \emph{Proc. {IEEE} {INFOCOM}}, Phoenix, AZ,
  USA, April 2008.

\bibitem{CaireTB99j}
G.~Caire, G.~Taricco, and E.~Biglieri, ``Optimum power control over fading
  channelsl,'' \emph{{IEEE} Trans. Info. Theory}, vol.~45, no.~5, pp.
  1468--1489, 1999.

\bibitem{LiG00j2}
L.~Li and A.~Goldsmith, ``Capacity and optimal resource allocation for fading
  broadcast channels: Part {II}: outage capacity,'' \emph{{IEEE} Trans. Info.
  Theory}, vol.~47, no.~3, pp. 120--145, 2000.

\bibitem{LanemanTW04j}
J.~N. Laneman, D.~N.~C. Tse, and G.~W. Wornell, ``Cooperative diversity in
  wireless networks: efficient protocols and outage behavior,'' \emph{{IEEE}
  Trans. Info. Theory}, vol.~50, pp. 3062--3080, 2004.

\bibitem{QiuC99j}
X.~Qiu and K.~Chawla, ``On the performance of adaptive modulation in cellular
  systems,'' \emph{{IEEE} Trans. Commun.}, vol.~47, no.~6, pp. 884--895, June
  1999.

\bibitem{OymanS06c}
O.~Oyman and S.~Sandhu, ``Non-ergodic power-bandwidth tradeoff in linear
  multi-hop networks,'' in \emph{Proc. ISIT}, Seattle, Washington, USA, July
  2006.

\bibitem{BoydV04b}
S.~Boyd and L.~Vandenberghe, \emph{Convex Optimization}.\hskip 1em plus 0.5em
  minus 0.4em\relax Cambridge, United Kingdom: Cambridge Univ. Press, 2004.

\bibitem{TaoLZ06c}
M.~Tao, Y.~Liang, and F.~Zhang, ``Adaptive resource allocation for delay
  differentiated traffics in multiuser ofdm systems,'' accepted for publication
  in \emph{IEEE Trans. Wirel. Commun.}

\bibitem{YaoG05j}
Y.~Yao and G.~B. Giannakis, ``Energy-efficient scheduling for wireless sensor
  networks,'' \emph{{IEEE} Trans. Commun.}, vol.~53, no.~8, pp. 1333--1342,
  Aug. 2005.

\bibitem{GamalNPUZ02c}
A.~E. Gamal, C.~Nair, B.~Prabhakar, E.~Uysal-Biyikoglu, and S.~Zahedi,
  ``Energy-efficient scheduling of packet transmissions over wireless
  networks,'' in \emph{{IEEE} {INFOCOM}}, New York, USA, March 2002.

\bibitem{ErcegH01s}
V.~Erceg, K.~Hari, M.~Smith, and D.~{Baum \emph{et al}}, ``Channel models for
  fixed wireless applications,'' {IEEE} 802.16.3c-01/29r1, 23 Feb 2001.

\bibitem{ErcegGTP99j}
V.~Erceg, L.~Greenstein, S.~Tjandra, S.~Parkoff, A.~Gupta, B.~Kulic, A.~Julius,
  and R.~Jastrzab, ``An empirically based path loss model for wireless channels
  insuburban environments,'' \emph{{IEEE} J. Sel. Areas Commun.}, vol.~2,
  no.~11, pp. 1205--1211, 8-12 Nov. 1999.

\bibitem{SikoraLHCF04c}
M.~Sikora, J.~N. Laneman, M.~Haenggi, J.~D.~J.~Costello, and T.~E. Fuja, ``On
  the optimum number of hops in linear ad hoc networks,'' in \emph{Proc. {IEEE}
  Info. Theory Workshop}, San Antonio, Oct. 2004, pp. 165--169.

\end{thebibliography}

\newpage

%

\end{document}